\documentclass[twocolumn,prb,showpacs]{revtex4}% Physical Review B
\usepackage{graphicx}% Include figure files
\usepackage{dcolumn}% Align table columns on decimal point
\usepackage{bm}% bold math

\begin{document}
\title{Interband electron Raman scattering in a quantum wire in a
transverse magnetic field}
\author{F.M. Hashimzade\dag, T.G. Ismailov\ddag, B.H. Mehdiyev\dag}
\address{\dag Institute of Physics, Azerbaijan Academy
of Sciences, AZ 1143, Baku,
Azerbaijan,\\
\ddag Baku State University, AZ 1073/1, Baku, Azerbaijan}
\author{S. T. Pavlov\dag\ddag}
\address{\dag Facultad de Fisica de la UAZ, Apartado Postal C-580, 98060 Zacatecas, Zac., Mexico\\
\ddag P. N. Lebedev Physical Institute, Russian Academy of
Sciences, 119991 Moscow, Russia}

\begin{abstract}
Electron Raman scattering (ERS) is investigated in a parabolic semiconductor
quantum wire in a transverse magnetic field neglecting by phonon-assisted
transitions. The ERS cross-section is calculated as a function of a
frequency shift and magnetic field. The process involves an interband
electronic transition and an intraband transition between quantized
subbands. We analyze the differential cross-section for different scattering
configurations. We study selection rules for the processes. Some
singularities in the Raman spectra are found and interpreted. The scattering
spectrum shows density-of-states peaks and interband matrix elements
maximums and a strong resonance when scattered frequency equals to the
"hybrid" frequency or confinement frequency depending on the light
polarization. Numerical results are presented for a GaAs/AlGaAs quantum wire.
\end{abstract}
\pacs{78.67.Lt; 78.30.Fs} \maketitle

\section{Introduction}

Low-dimensional semiconductor systems, quantum wires in particular, attract
considerable attention, because of their novel physical properties and
application potential. In recent years a number of innovative techniques
have been developed to grow or to fabricate and to study experimentally a
variety of quantum wire structures having different geometries and
potentials. Many recent experimental and theoretical studies have been
performed on quantum wires subjected to a transverse magnetic field [1-7].
Electronic properties of quantum wells in a transverse magnetic field have
been investigated in [8-9]. The subband dispersion and magnetoabsorption
have been studied for rectangular quantum wires in [10].

A magnetic field perpendicular to the wire axis (a "free electron"
direction) can change significantly the electronic states of a semiconductor
quantum wire.

Electron Raman scattering seems to be a useful technique providing a direct
information on the energy band structure and optical properties of
investigated systems [11-13]. In particular, the electronic structure of
semiconductor materials and nanostructures can be thoroughly investigated
considering different polarizations for the incident and emitted radiation
[14]. The differential cross-section, in general case, usually shows
singularities related to interband and intraband transitions. This latter
result strongly depends on the scattering configurations: the structure of
singularities varies when photon polarizations change. This feature of the
ERS allows to determine the subband structure of the system by a direct
inspection of singularity positions in the spectra. For bulk semiconductors
the ERS has been studied in the presence of external magnetic and electric
fields [15-17]. In the case of a quantum well preliminary results were
reported in [18].

Raman scattering in low-dimensional semiconductor systems has been
the subject of many theoretical and experimental investigations
[19,20]. Interband ERS processes can be qualitatively described in
the following way: absorption of a photon of the incident
radiation field creates a virtual electron-hole pair (EHP) in an
intermediate crystal state by means of an electron interband
transition involving the crystal valence and conduction bands. An
electron in the conduction band is subject to a second intraband
transition with emission of a secondary radiation photon.
Therefore, in the final state we have a real EHP in the crystal
and a photon of the secondary radiation. The influence of external
fields on such processes for bulk semiconductors is investigated
in [16, 17].

In this work we present a systematic study of the interband ERS in
a direct band gap semiconducting parabolic quantum wire in a
transverse magnetic field. In these systems due to electron
confinement and magnetic field the conduction (valence) band is
split in a subband system and transitions between them determine
ERS processes. Numerical results for the ERS differential
cross-section are presented for a GaAs/AlGaAs quantum wire. This
artical is organized as follows. In Section II the energy spectrum
and wave functions for a quantum wire with parabolic confinement
potential in a transverse magnetic field are given. In Section III
we present the general relations needed for our calculations of
the ERS differential cross-section. Section IV is devoted to
calculations of ERS differential cross-sections. Finally, Section
V is concerned with the discussion of the obtained results.

\section{Wave functions and energy spectrum}

We consider a quantum wire aligned along the $y$ axis with a
transverse magnetic field $\mathbf{H}=\mathbf{H}(0,0,H)$ applied
along the $z$ axis.
\begin{figure}
\includegraphics[]{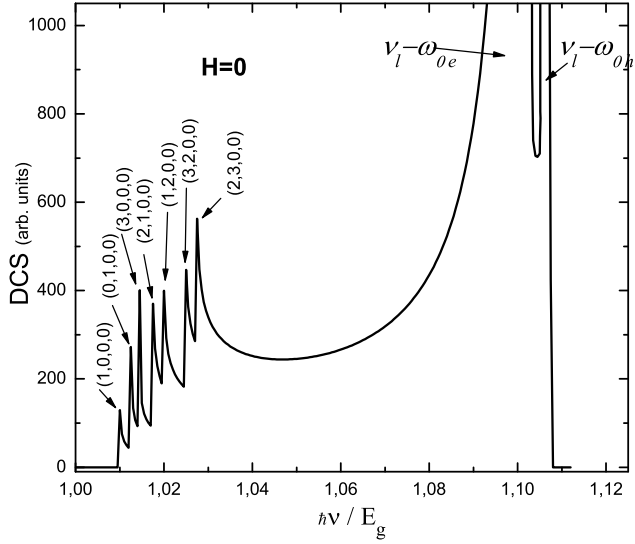}% Here is how to import EPS art
\caption[*]{\label{fg1.eps}The Raman spectra of the parabolic
quantum wire in the $X$ scattering configuration for H=0. The
diameter $d$ of the quantum wire is  2000 A. The incident
radiation frequency  $\hbar\nu_0= 1.68$ eV. The positions of the
singularities are defined by Eqs. (42) and (43). Resonant
electron-hole transitions are indicated by $N_{1h}, N_{1e},
N_{2h}, N_{2e}$. }
\end{figure}
The quantum wire is characterized by parabolic confinements in the
$(x,z)$ plane. The effective mass Schr\"{o}edinger equation for an
electron in a conduction band is
\begin{eqnarray}
\label{1}\left[{1\over 2m_e}\left({\bf p}+{e\over c}{\bf
A}\right)^2+{1\over 2}m_e\omega^2_{0e}(x^2+z^2)\right]\psi_e(x, y,
z)\nonumber\\=E_e\psi_e(x, y, z),
\end{eqnarray}
where $\mathbf{A}=\mathbf{A}(0,xH,0)$ is the vector potential in
the Landau gauge; $\omega _{0e}$ characterizes the parabolic
potential of a quantum wire for electrons in a conduction band;
$m_{e},-e$ are the electron effective mass and charge,
respectively. We look for the solution in the form
$$\psi_e(x, y, z)=\varphi(x)\eta(z)e^{ip_{y,e}y/\hbar},$$
where $p_{y,e}=\hbar k_{y,e}$ is the quasi-momentum of an electron.

Shifting the origin of coordinates and separating the variables in
the usual way we obtain the eigenfunctions and eigenvalues of the
Schr\"oedinger equation (1)
\begin{equation}
\label{2}
\psi_{N_{1e}, N_{2e}, k_{y,e}}=\varphi_{N_{1e}}\left({\frac{x-x_{0e}}{{%
\tilde L}_e}}\right) \eta_{N_{2e}}\left({\frac{z}{L_e}}\right)e^{ik_{y,e}y},
\end{equation}
\begin{eqnarray}
\label{3}
E_e&=&(N_{1e}+1/2)\hbar{\tilde\omega}_e+(N_{2e}+1/2)\hbar\omega_{0e}\nonumber\\
&+& {\frac{%
\hbar^2k_{y,e}^2}{2m_e}}\left({\frac{\omega_{0e}
}{{\tilde\omega}_e}} \right)^2.
\end{eqnarray}

The wave functions and energy eigenvalues for electrons in valence band are
as follows
\begin{equation}
\label{4} \psi_{N_{1h}, N_{2h},
k_{y,h}}=\varphi_{N_{1h}}\left({x-x_{0h}\over {\tilde L}_e}\right)
\eta_{N_{2h}}\left({z\over L_h}\right)e^{ik_{y,h}y},
\end{equation}
\begin{eqnarray}
\label{5}E_h=-E_g-(N_{1h}+1/2)\hbar{\tilde\omega}_h+(N_{2h}+1/2)\hbar\omega_{0h}\nonumber\\-
{\hbar^2k_{y,h}^2\over2m_h}\left({\omega_{0h}
\over{\tilde\omega}_h}\right)^2,
\end{eqnarray}
where $E_g$ is the energy gap between the valence and conduction bands in
absence of an external magnetic field, $\omega_{0h}$ is the oscillator
frequency of the parabolic potential for electrons in the valence band. In
Eqs. (2) - (5)
\begin{equation}
\label{6}
{\tilde\omega_{e,h}}=\sqrt{\omega^2_{0e(h)}+\omega^2_{e(h)}}
\end{equation}
is the "hybrid" frequency. The subscripts $e$ and $h$ denote
conduction and valence band, respectively.
\begin{equation}
\label{7}
\omega_{e(h)}={\frac{eH}{m_{e(h)}c}}
\end{equation}
is the cyclotron frequency, $m_{e(h)}$ is the effective mass,
\begin{equation}
\label{8}
x_{0e(h)}={\frac{\hbar\omega_{e(h)}}{m_{e(h)}{\tilde \omega}_{e(h)}^2}}\cdot
k_{ye(h)}
\end{equation}
is the oscillator centre.

The full energy spectrum in (2) - (5) is governed by quantum numbers $%
N_{1e(h)}$ , $N_{2e(h)}$ and $k_{ye(h)}$.

\begin{eqnarray}
\label{9} \varphi_{N_{1e(h)}}\left({x-x_{0e(h)}\over {\tilde
L}_{e(h)}}\right)=\left({1\over\pi{\tilde
L}_{e(h)}^2}\right)^{1/4}{1\over\sqrt{2^{N_{1e(h)}}N_{1e(h)}!}}\nonumber\\
\times\exp\left(-{(x-x_{0h})^2\over 2{\tilde
L}_e^2}\right)H_{N_{1e(h)}}\left({x-x_{0h}\over {\tilde
L}_{e(h)}}\right),
\end{eqnarray}
\begin{eqnarray}
\label{10}\eta_{N_{2e(h)}}\left({z\over
L_{e(h)}}\right)=\left({1\over\pi
L_{e(h)}^2}\right)^{1/4}{1\over\sqrt{2^{N_{2e(h)}}N_{2e(h)}!}}\nonumber\\
\times\exp\left(-{z^2\over
 2L_{e(h)}^2}\right)H_{N_{2e(h)}}\left({z\over L_{e(h)}}\right),
\end{eqnarray}
where parameters
\begin{equation}
\label{11}
{\tilde L}_{e(h)}=\sqrt{{\frac{\hbar}{m_{e(h)}{\tilde \omega}_{e(h)}}}};~~~
L_{e(h)}=\sqrt{{\frac{\hbar}{m_{e(h)} \omega_{0e(h)}}}}
\end{equation}
are the units of length; $H_n(\xi)$ is the Hermitian polynomial.

\section{Preliminary relations}

The general expression for the ERS differential cross-section is given by
[16,18]

\begin{equation}
\label{12}
\frac{d^{2}\sigma }{d\Omega d\nu _{s}}=\frac{V^{2}\nu _{s}^{2}n(\nu _{s})}{%
8\pi ^{3}c^{4}n(\nu _{l})}W(\nu _{s},\mathbf{e}_{s})
\end{equation}
where $c$ is the light velocity in vacuum, $n(\nu )$ is refraction
index as a function of the radiation frequency, $\mathbf{e}_{s}$
the (unit) polarization vector for the secondary radiation field,
$V$ is the normalization volume, $\nu _{s}$ is the secondary
radiation frequency, $\nu
_{l}$ is the frequency of the incident radiation. $W(\nu _{s},\mathbf{e}%
_{s}) $ is the transition rate calculated according to
\begin{equation}
\label{13} W(\nu _{s}, {\bf e}_{s})=\frac{2\pi }{\hbar }\sum_f
\left| M_{e}+M_{h}\right|^{2}\delta (E_{f}-E_{i}),
\end{equation}
where
\begin{eqnarray}
\label{14}
M_{j}=\sum_a\frac{\left\langle f\left\vert \hat{H}%
_{js}\right\vert a\right\rangle \left\langle a\left\vert \hat{H}%
_{l}\right\vert i\right\rangle }{E_{i}-E_{a}}\nonumber\\
+\sum_b\frac{%
\left\langle f\left\vert \hat{H}_{l}\right\vert b\right\rangle
\left\langle b\left\vert \hat{H}_{js}\right\vert i\right\rangle
}{E_{i}-E_{b}}.
\end{eqnarray}

In (14) $j=e,h$ are for the cases of electrons or holes, respectively, $%
|i\rangle $ and $|f\rangle $ denote initial and final states of the system
with their corresponding energies $E_{i}$ and $E_{f}$. $|a\rangle $ and $%
|b\rangle $ are intermediate states with energies $E_{a}$ and $E_{b}$.

The operator $\hat{H}_{l}$ is of the form
\begin{equation}
\label{15}
\hat{H}_{l}=\frac{\left\vert e\right\vert }{m_{0}}\sqrt{\frac{2\pi \hbar }{%
V\nu _{l}}}\mathbf{e}_{l}\cdot \mathbf{\hat{p},\ \ \ \ \ \hat{p}=-}i\mathbf{%
\hbar \nabla },
\end{equation}
where $m_{0}$ is the free electron mass. This operator describes
the interaction with the incident radiation field in the dipol
approximation. The interaction with the secondary -radiation field
is described by the operator
\begin{equation}
\label{16}
\hat{H}_{js}=\frac{\left\vert e\right\vert }{m_{j}}\sqrt{\frac{2\pi \hbar }{%
V\nu _{s}}}\mathbf{e}_{s}\cdot \mathbf{\hat{p},\ \ \ }\
j\mathbf{=}e,h.
\end{equation}

This Hamiltonian describes the photon emission by the electron (hole) after
transitions between conduction (valence) subbands of the system. In (14) the
intermediate states $|a\rangle $ represent an EHP in a virtual state (after
absorption of the incident photon), while the states $|b\rangle $ are
related to the \textquotedblleft interference diagrams\textquotedblright\
[16,18]. This latter term involves a negligible contribution whenever the
energy gap $E_{g}$ is large enough (for instanse, this is the case for $GaAs$%
) and will not be considered in the present work.

We established that the proscesses of ERS given in following are possible:

\subparagraph{a) The interband ERS process with the intermediate state in
the conduction band.}

First, an incident light quantum is being absorbed creating an electron
-hole pair between the state $(N_{1h},N_{2h})$ in the valence band and the
state $(N_{1e}^{\prime },N_{2e}^{\prime })$ in the conduction band. Second,
a scattered photon is emitted due to an electronic transition from the state
$(N_{1e}^{\prime },N_{2e}^{\prime })$ to the state $(N_{1e},N_{2e})$ in the
conduction band. The Raman shift $\hbar \nu =\hbar (\nu _{l}-\nu _{s})$ is
equal to the excitation energy of the electron-hole pair created in the
scattering process.

\subparagraph{b) The interband ERS process with the intermediate state in
the valence band.}

Two electrons take part in the process. After the absorption of incident
photon the first electron from the state $(N_{1h}^{^{\prime
}},N_{2h}^{^{\prime }})$ in valence band is lifted to the state $%
(N_{1e},N_{2e})$. The second electron from the state $(N_{1h},N_{2h})$ falls
to the vacant state in the $(N_{1h}^{^{\prime }},N_{2h}^{^{\prime }})$
subband. The real transition corresponds to a transition from the state $%
(N_{1h},N_{2h})$ to the state $(N_{1e},N_{2e})$.

In the initial state $|i\rangle $ we have an incident radiation
photon with frequency $\nu _{l}$, while the conduction band is
empty and the valence band completely occupied by electrons. We
neglect by all the transitions assisted by phonons.
\begin{figure}
\includegraphics[]{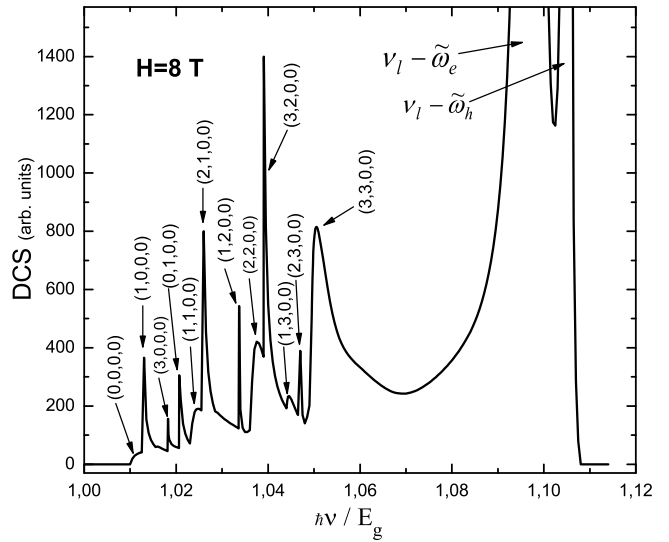}% Here is how to import EPS art
\caption[*]{\label{fg2.eps}Same that in Fig.1 for $ H=8T$.}
\end{figure}

The initial state energy is
\begin{equation}
\label{17} E_{i}=\hbar \nu _{l}.
\end{equation}

The final state of the process consists of an EHP in a real state and a
scattered light with energy $\hbar \nu _{s}$. Thus,
\begin{eqnarray}
\label{18}
E_{f}=\hbar \nu _{s}+E_{N_{1h}}+E_{N_{2h}}+E_{N_{1e}}+E_{N_{2e}}+E_{g}\nonumber\\
+\frac{%
\hbar ^{2}k_{ye}^{2}}{2m_{e}}\left( \frac{\omega _{0e}}{\tilde{\omega}_{e}}%
\right) ^{2}+\frac{\hbar ^{2}k_{yh}^{2}}{2m_{h}}\left( \frac{\omega _{0h}}{%
\tilde{\omega}_{h}}\right) ^{2},
\end{eqnarray}
where
\begin{eqnarray}
\label{19}
E_{N_{1e(h)}} &=&(N_{1e(h)}+1/2)\hbar {\tilde{\omega}}_{e(h)}, \nonumber\\
E_{N_{2e(h)}} &=&(N_{2e(h)}+1/2)\hbar \omega _{0e(h)}.
\end{eqnarray}

For the electron intermediate states $|a\rangle $ the energies $E_{a}$ are
easily obtained from the above discussion.
\begin{eqnarray}
\label{20}
E_{i}-E_{a}=-E_{g}-E_{N_{1h}}-E_{N_{2h}}-E_{N_{1e}^{^{\prime
}}}-E_{N_{2e}^{^{\prime }}}\nonumber\\-\frac{\hbar ^{2}k_{ye}^{2}}{2m_{e}}\left( \frac{%
\omega _{0e}}{\tilde{\omega}_{e}}\right) ^{2}
-\frac{\hbar ^{2}k_{yh}^{2}}{%
2m_{h}}\left( \frac{\omega _{0h}}{\tilde{\omega}_{h}}\right)
^{2}+\hbar \nu _{l}.
\end{eqnarray}

Similar expressions can be written for the hole intermediate state energies.

\section{Calculation of the Raman scattering cross section}

The matrix elements of the intraband transitions may be written as
\begin{eqnarray}
\label{21} &&\frac{\left| e\right| }{m_{e}}\sqrt{\frac{2\pi \hbar
}{V\nu _{s}}}\langle
N_{1e},N_{2e},k_{ye}|\mathbf{e_{s}}\mathbf{p}|N_{1e}^{\prime
}N_{2e}^{\prime },k_{ye}^{\prime }\rangle
=\nonumber\\
&=&\frac{\left| e\right| }{m_{e}}\sqrt{\frac{2\pi \hbar }{V\nu
_{s}}}\left[ \langle
N_{1e},N_{2e},k_{ye}|e_{sx}p_{x}|N_{1e}^{\prime }N_{2e}^{\prime
},k_{ye}^{\prime }\rangle +\right.  \nonumber\\
&+&\langle N_{1e},N_{2e},k_{ye}|e_{sy}p_{y}|N_{1e}^{\prime
}N_{2e}^{\prime
},k_{ye}^{\prime }\rangle +  \nonumber\\
 &+&\left.\langle N_{1e},N_{2e},k_{ye}|e_{sz}p_{z}|N_{1e}^{\prime
}N_{2e}^{\prime },k_{ye}^{\prime }\rangle \right] ,
\end{eqnarray}
where
\begin{eqnarray}
\label{22} &&\langle
N_{1e},N_{2e},k_{ye}|e_{sx}p_{x}|N_{1e}^{\prime }N_{2e}^{\prime
},k_{ye}^{\prime }\rangle =-{\frac{i\hbar
}{{\tilde{L}}_{e}}}e_{sx}\delta
_{N_{2e}^{\prime },N_{2e}}  \nonumber\\
&\times& \left[ \sqrt{{\frac{N_{1e}}{2}}}\delta _{N_{1e}^{\prime },N_{1e}-1}-%
\sqrt{{\frac{N_{1e}+1}{2}}}\delta _{N_{1e}^{\prime
},N_{1e}+1}\right]\nonumber\\
&\times& \delta _{k_{ye},k_{ye}^{\prime }},
\end{eqnarray}
\begin{eqnarray}
\label{23} \langle N_{1e},N_{2e},k_{ye}|e_{sy}p_{y}|N_{1e}^{\prime
}N_{2e}^{\prime
},k_{ye}^{\prime }\rangle =  \nonumber\\
=\hbar k_{ye}^{\prime }e_{sy}\delta _{N_{2e}^{\prime
},N_{2e}}\delta _{N_{2e}^{\prime },N_{2e}}\delta
_{k_{ye},k_{ye}^{\prime }},
\end{eqnarray}
\begin{eqnarray}
\label{24}
 &&\langle N_{1e},N_{2e},k_{ye}|e_{sz}p_{z}|N_{1e}^{\prime
}N_{2e}^{\prime },k_{ye}^{\prime }\rangle =-{\frac{i\hbar
}{L_{e}}}e_{sz}\delta
_{N_{1e}^{\prime },N_{1e}} \nonumber\\
&\times& \left[ \sqrt{{\frac{N_{2e}}{2}}}\delta _{N_{2e}^{\prime },N_{2e}-1}-%
\sqrt{{\frac{N_{2e}+1}{2}}}\delta _{N_{2e}^{\prime
},N_{2e}+1}\right]\nonumber\\
&\times& \delta _{k_{ye},k_{ye}^{\prime }}.
\end{eqnarray}

A similar expression can be written for the interband ERS process with the
intermediate state in the valence band
\begin{eqnarray}
\label{25}
&&\frac{\left\vert e\right\vert }{m_{h}}\sqrt{\frac{2\pi \hbar }{V\nu _{s}}}%
\langle N_{1h}^{^{\prime }},N_{2h}^{^{\prime }},k_{yh}^{^{\prime }}|\mathbf{%
e_{s}}\mathbf{p}|N_{1h,}N_{2h},k_{yh}\rangle
=\nonumber\\
&=&\frac{\left\vert e\right\vert }{m_{h}}\sqrt{\frac{2\pi \hbar
}{V\nu _{s}}}\left[ \langle N_{1h}^{^{\prime }},N_{2h}^{^{\prime
}},k_{yh}^{^{\prime
}}|e_{sx}p_{x}|N_{1h,}N_{2h},k_{yh}\rangle \right.\nonumber\\
&+&\langle N_{1h}^{^{\prime }},N_{2h}^{^{\prime
}},k_{yh}^{^{\prime
}}|e_{sy}p_{y}|N_{1h,}N_{2h},k_{yh}\rangle\nonumber\\
&+&\left.\langle N_{1h}^{^{\prime }},N_{2h}^{^{\prime
}},k_{yh}^{^{\prime }}|e_{sz}p_{z}|N_{1h,}N_{2h},k_{yh}\rangle
\right] ,
\end{eqnarray}
where
\begin{eqnarray}
\label{26} &&\langle N_{1h}^{^{\prime }},N_{2h}^{^{\prime
}},k_{yh}^{^{\prime }}|e_{sx}p_{x}|N_{1h,}N_{2h},k_{yh}\rangle
=\nonumber\\
&=&-{\frac{i\hbar }{{\tilde{L}}%
_{h}}}e_{sx}\delta _{N_{2h},N_{2h}^{^{\prime }}}\nonumber\\
&\times& \left[ \sqrt{{\frac{N_{1h}+1}{2}}}\delta
_{N_{1h},N_{1h}^{^{\prime }}-1}-\sqrt{{\frac{N_{1h}}{2}}}\delta
_{N_{1h},N_{1h}^{^{\prime }}+1}\right]\nonumber\\
&\times&\delta _{k_{yh},k_{yh}^{\prime }},
\end{eqnarray}
\begin{eqnarray}
\label{27} \langle N_{1h}^{^{\prime }},N_{2h}^{^{\prime
}},k_{yh}^{^{\prime
}}|e_{sy}p_{y}|N_{1h,}N_{2h},k_{yh}\rangle = \nonumber\\
=\hbar k_{yh}e_{sy}\delta _{N_{2h},N_{2h}^{^{\prime }}}\delta
_{N_{1h},N_{1h}^{^{\prime }}}\delta _{k_{yh},k_{yh}^{\prime }},
\end{eqnarray}
\begin{eqnarray}
\label{28} &&\langle N_{1h}^{^{\prime }},N_{2h}^{^{\prime
}},k_{yh}^{^{\prime }}|e_{sz}p_{z}|N_{1h,}N_{2h},k_{yh}\rangle
=\nonumber\\
&\times&-{\frac{i\hbar }{L_{h}}}%
e_{sz}\delta _{N_{1h},N_{1h}^{^{\prime }}}\nonumber\\
&\times& \left[ \sqrt{{\frac{N_{2h}+1}{2}}}\delta
_{N_{2h},N_{2h}^{^{\prime }}-1}-\sqrt{{\frac{N_{2h}}{2}}}\delta
_{N_{2h},N_{2h}^{^{\prime }}+1}\right]\nonumber\\
&\times& \delta _{k_{yh},k_{yh}^{\prime }}.
\end{eqnarray}
\begin{figure}
\includegraphics[]{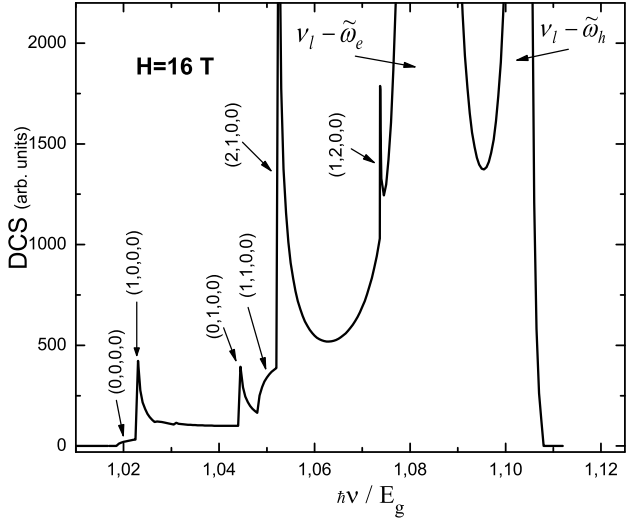}% Here is how to import EPS art
\caption[*]{\label{fg3.eps}Same that in Fig.1 for $ H=16T $.}
\end{figure}

If we consider allowed electron transitions between conduction and valence
bands, the interband matrix element in the envelope function approximation,
may be written as
\begin{eqnarray}
\label{29}
&&\langle a|\hat{H}_{l}|i\rangle =\frac{\left\vert e\right\vert }{m_{0}}\sqrt{%
\frac{2\pi \hbar }{V\nu
_{l}}}(\mathbf{p}_{cv}\mathbf{e}_{l})\nonumber\\
&\times&\left\{
\begin{array}{c}
I_{N_{1h},N_{1e}^{\prime }}(k_{y})J_{N_{2h},N_{2e}^{\prime }}\delta
_{k_{yh},k_{ye}^{\prime }},\ \ \ \ j=e;\ \  \\
I_{N_{1h}^{^{\prime }},N_{1e}}(k_{y})J_{N_{2h}^{^{\prime
}},N_{2e}}\delta _{k_{yh}^{^{\prime }},k_{ye}},\ \ \ j=h,\ \
\end{array}
\right. \ \
\end{eqnarray}
where $\mathbf{p}_{cv}$ is the momentum matrix element between the valence
and conduction bands (evaluated at $\mathbf{k}=0$ ).

We find that the matrix elements (22)-(29) vanish unless the following
selection rule is satisfied
\begin{equation}
\label{30} k_{ye}=k_{yh}=k_{ye}^{\prime }=k_{yh}^{\prime }=k_{y}.
\end{equation}

The EHP does not change its total momentum during absorption or
emission of a photon ( a photon momentum is neglected). It may be
obtained that
\begin{eqnarray}
\label{31}  &&I_{N_{1h},N_{1e}^{\prime }}(k_{y})=\left(
{\frac{1}{\pi }}\right) ^{1/2}\left(
{\frac{1}{{\tilde{L}}_{h}{\tilde{L}}_{e}}}\right) ^{1/2}
\nonumber\\
&\times &{\frac{N_{1h}!N_{1e}^{\prime
}!}{\sqrt{2^{N_{1h}+N_{1e}^{\prime
}}N_{1h}!N_{1e}^{\prime }!}}}  \nonumber\\
&\times &\sum_{k=0}^{[N_{1h}/2]}\sum_{k=0}^{[N_{1e}^{\prime }/2]}{\frac{%
(-1)^{k+j}2^{N_{1h}+N_{1e}^{\prime }-2k-2j}}{k!j!(N_{1h}-2k)!(N_{1e}^{\prime
}-2j)!}}  \nonumber\\
&\times &\left( \frac{1}{{\tilde{L}}_{e}}\right) ^{N_{1e}^{\prime
}-2j}\left( \frac{1}{{\tilde{L}}_{h}}\right) ^{N_{1h}-2k}\sum_{\mu
=0}^{N_{1h}-2k}{\frac{(N_{1h}-2k)!}{\mu !(N_{1h}-2k-\mu )!}}  \nonumber\\
&\times &(x_{0e}-x_{0h})^{N_{1h}-2k-\mu }\sum_{\nu =0}^{N_{1e}^{\prime
}-2j+\mu }{\frac{(N_{1e}^{\prime }-2j+\mu )!}{\nu !(N_{1e}^{\prime }-2j+\mu
-\nu )!}}  \nonumber\\
&\times &[(-1)^{\nu }+1]\exp \left( \frac{-{(x_{0e}-x_{0h})^{2}}}{2({\tilde{L%
}}_{e}^{2}+{\tilde{L}}_{h}^{2})}\right)\nonumber\\
&\times&{\frac{1}{2}}\left( {\frac{{\tilde{L}}_{e}^{2}+{\tilde{L}}_{h}^{2}}{%
2{\tilde{L}}_{e}^{2}{\tilde{L}}_{h}^{2}}}\right) ^{-(\nu +1)/2}  \nonumber\\
&\times &\Gamma \left( {\frac{\nu +1}{2}}\right) \left( -{\frac{{\tilde{L}}%
_{e}^{2}(x_{0e}-x_{0h})}{{\tilde{L}}_{e}^{2}+{\tilde{L}}_{h}^{2}}}\right)
^{N_{1e}^{^{\prime }}-2j+\mu -\nu }
\end{eqnarray}
and
\begin{eqnarray}
\label{32}
&&J_{N_{2h},N_{2e}^{\prime }}=\left( {\frac{1}{\pi }}\right) ^{1/2}\left( {%
\frac{1}{L_{h}L_{e}}}\right) ^{1/2}\sqrt{{\frac{N_{2h}!N_{2e}^{\prime }!}{%
2^{N_{2h}+N_{2e}^{\prime }}}}}  \nonumber\\
&\times &\sum_{\alpha =0}^{[N_{2e}^{\prime }/2]}\sum_{\beta =0}^{[N_{2h}/2]}{%
\frac{(-1)^{\alpha +\beta }2^{N_{2e}^{\prime }-2\alpha +N_{2h}-2\beta }}{%
\alpha !\beta !(N_{2e}^{\prime }-2\alpha )!(N_{2h}-2\beta )!}}  \nonumber\\
&\times &\left( \frac{1}{L_{e}}\right) ^{N_{2e}^{\prime }-2\alpha }\left(
\frac{1}{L_{h}}\right) ^{N_{2h}-2\beta }[(-1)^{N_{2e}^{\prime
}+N_{2h}-2\alpha -2\beta }+1]  \nonumber\\
&\times &{\frac{\Gamma \left( {\frac{N_{2e}^{\prime }+N_{2h}-2\alpha -2\beta
+1}{2}}\right) }{2\left( {\frac{\sqrt{L_{e}^{2}+L_{h}^{2}}}{\sqrt{2}%
L_{h}L_{e}}}\right) ^{N_{2e}^{\prime }+N_{2h}-2\alpha -2\beta
+1}}}.
\end{eqnarray}

An expression analogous to Eqs. (31) and (32) holds for $I_{N_{1h}^{^{\prime
}},N_{1e}}(k_{y})$ and $J_{N_{2h}^{^{\prime }},N_{2e}}$ after making the
replacements $N_{1h}\rightarrow N_{1h}^{^{\prime }}$ and $N_{2h}\rightarrow
N_{2h}^{^{\prime }}$.

 Performing the summation over $k_{y}$ in (13) we obtain the ERS
differential cross section:
\begin{eqnarray}
\label{33} \frac{d^{2}\sigma }{d\Omega d\nu _{s}}=\left(
\frac{d^{2}\sigma }{d\Omega
d\nu _{s}}\right) _{e_{sx}}+\left( \frac{d^{2}\sigma }{d\Omega d\nu _{s}}%
\right) _{e_{sy}}\nonumber\\
+\left( \frac{d^{2}\sigma }{d\Omega
d\nu _{s}}\right) _{e_{sz}},
\end{eqnarray}
where
\begin{eqnarray}
\label{34} &&\left( \frac{d^{2}\sigma }{d\Omega d\nu_{s}}\right)
_{e_{sx}} =\frac{
\sigma _{0}}{\tilde{L}_{e}^{2}}{\frac{\nu _{l}-\nu }{\nu _{l}}}\nonumber\\
&\times&\sum_{N_{1e},N_{2e},N_{1h},N_{2h}}\Big[\sum_{N_{1e}^{\prime
},N_{2e}^{\prime },N_{1h}^{\prime },N_{2h}^{\prime
}}\nonumber\\
&\times&\left\{ \frac{\beta }{A(\nu )}\delta
_{N_{2e},N_{2e}^{\prime }}\right. \nonumber\\
&\times& \left( \sqrt{N_{1e}/2}\cdot \delta _{N_{1e}^{\prime },N_{1e}-1}-%
\sqrt{(N_{1e}+1)/2}\cdot \delta _{N_{1e}^{\prime
},N_{1e}+1}\right)
\nonumber\\
&\times& I_{N_{1h},N_{1e}^{\prime }}(k_{y}(\nu
))J_{N_{2h},N_{2e}^{\prime }}\nonumber\\
&+&\frac{\gamma (H)}{B\left( \nu \right) }\delta
_{N_{2h},N_{2h}^{^{\prime }}}I_{N_{1h}^{^{\prime
}},N_{1e}}(k_{y}(\nu ))J_{N_{2h}^{^{\prime
}},N_{2e}} \nonumber\\
&\times& \left( \sqrt{{\frac{N_{1h}+1}{2}}}\delta
_{N_{1h},N_{1h}^{^{\prime
}}-1}-\sqrt{{\frac{N_{1h}}{2}}}\delta _{N_{1h},N_{1h}^{^{\prime }}+1}\right) %
\Big]^{2}\nonumber\\
&\times& \left( {\frac{E_{g}}{\hbar \nu
-E_{N_{1e}}-E_{N_{1h}}-E_{N_{2e}}-E_{N_{2h}}-E_{g}}}\right)
^{1/2}\nonumber\\
&\times&\left\vert \mathbf{e}_{s}\cdot \mathbf{X}\right\vert ^{2},
\end{eqnarray}
and
\begin{eqnarray}
\label{35} &&\left( \frac{d^{2}\sigma }{d\Omega d\nu _{s}}\right)
_{e_{sy}} =\frac{
\sigma _{0}}{L_{e}^{2}}{\frac{\nu _{l}-\nu }{\nu _{l}}}\nonumber\\
&\times&\sum_{N_{1e},N_{2e},N_{1h},N_{2h}}
\Big[\sum_{N_{1e}^{\prime },N_{2e}^{\prime
},N_{1h}^{\prime },N_{2h}^{\prime }}\nonumber\\
&\times&\left\{ \frac{\beta }{A(\nu )}\right.
I_{N_{1h},N_{1e}^{\prime }}(k_{y}(\nu ))J_{N_{2h},N_{2e}^{\prime
}}\delta _{N_{2e},N_{2e}^{\prime
}}\delta _{N_{1e},N_{1e}^{\prime }} \nonumber\\
&+&\left.\frac{1}{B\left( \nu \right) }I_{N_{1h}^{^{\prime
}},N_{1e}}(k_{y}(\nu ))J_{N_{2h}^{^{\prime }},N_{2e}}\delta
_{N_{2h},N_{2h}^{^{\prime }}}\delta _{N_{1h},N_{1h}^{^{\prime
}}}\right\}
\Big]^{2} \nonumber\\
&\times& \left( {\frac{E_{g}}{\hbar \nu
-E_{N_{1e}}-E_{N_{1h}}-E_{N_{2e}}-E_{N_{2h}}-E_{g}}}\right)
^{1/2}\nonumber\\
&\times&\left( k_{y}(\nu )L_{e}\right) ^{2}\left\vert
\mathbf{e}_{s}\cdot \mathbf{Y}\right\vert ^{2},
\end{eqnarray}
\begin{eqnarray}
\label{36} &&\left( \frac{d^{2}\sigma }{d\Omega d\nu _{s}}\right)
_{e_{sz}} =\frac{ \sigma _{0}}{L_{e}}{\frac{\nu _{l}-\nu }{\nu
_{l}}}\nonumber\\
&\times& \sum_{N_{1e},N_{2e},N_{1h},N_{2h}}
\Big[\sum_{N_{1e}^{\prime },N_{2e}^{\prime },N_{1h}^{\prime
},N_{2h}^{\prime }}\left\{ \frac{\beta }{A(\nu )}\delta
_{N_{2e},N_{2e}^{\prime }}\right. \nonumber\\
&\times& \left( \sqrt{N_{2e}/2}\cdot \delta _{N_{2e}^{\prime
},N_{2e}-1}- \sqrt{(N_{2e}+1)/2}\cdot \delta _{N_{2e}^{\prime
},N_{2e}+1}\right)
\nonumber\\
&\times& I_{N_{1h},N_{1e}^{\prime }}(k_{y}(\nu
))J_{N_{2h},N_{2e}^{\prime }}\nonumber\\
&+& \frac{\gamma }{B\left( \nu \right) }\delta
_{N_{1h},N_{1h}^{^{\prime }}}I_{N_{1h}^{^{\prime
}},N_{1e}}(k_{y}(\nu ))J_{N_{2h}^{^{\prime
}},N_{2e}} \nonumber\\
&\times& \left( \sqrt{{\frac{N_{2h}+1}{2}}}\delta
_{N_{2h},N_{2h}^{^{\prime }}-1}-\sqrt{{\frac{N_{2h}}{2}}}\delta
_{N_{2h},N_{2h}^{^{\prime }}+1}\right)
\Big]^{2}\nonumber\\
&\times& \left( {\frac{E_{g}}{\hbar \nu
-E_{N_{1e}}-E_{N_{1h}}-E_{N_{2e}}-E_{N_{2h}}-E_{g}}}\right)
^{1/2}\nonumber\\
&\times&\left\vert \mathbf{e}_{s}\cdot \mathbf{Z}\right\vert ^{2},
\end{eqnarray}
where
\begin{eqnarray}
\label{37} \sigma _{0}&=&{e^{4}L_{y}|{\bf p}_{cv}{\bf
e}_{l}|^{2}\hbar^{2}n(\nu _{s})\over\sqrt{2}\pi
m_{0}^{2}m_{h}^{2}E_{g}^{5/2}n(\nu
_{l})c^{4}}\nonumber\\
&\times&{1\over\sqrt{{1\over m_e}\left( {\omega
_{0e}\over{\tilde{\omega}}_e}\right)^{2}+{1\over m_h}\left(
{\omega _{0h}\over{\tilde{ \omega}}_{h}}\right)^2}};
\end{eqnarray}
\begin{equation}
\label{38} A(\nu )={\frac{\hbar }{E_{g}}}[\nu -\nu
_{l}+(N_{2e}-N_{2e}^{\prime })\omega _{0e}+(N_{1e}-N_{1e}^{\prime
}){\tilde{\omega}}_{e}],
\end{equation}
\begin{equation}
\label{39} B(\nu )={\frac{\hbar }{E_{g}}}[(N_{2h}^{^{\prime
}}-N_{2h})\omega _{0h}+(N_{1h}^{^{\prime
}}-N_{1h}){\tilde{\omega}}_{h}-\nu +\nu _{l}],
\end{equation}
\begin{equation}
\label{40}
\beta =\frac{m_{h}}{m_{e}},\ \ \gamma (H)=\frac{\tilde{L}_{e}}{\tilde{L}_{h}}%
,\ \ \ \gamma =\frac{L_{e}}{L_{h}}
\end{equation}
and $k_{y}(\nu )$ is the root of the delta function argument
\begin{equation}
\label{41} |k_{y}(\nu )|={\frac{\sqrt{2}}{\hbar
}}\sqrt{\frac{\hbar \nu
-E_{g}-E_{N_{1e}}-E_{N_{1h}}-E_{N_{2e}}-E_{N_{2h}}}{{\frac{1}{m_{e}}}\left( {%
\frac{\omega _{0e}}{{\tilde{\omega}}_{e}}}\right) ^{2}+{\frac{1}{m_{h}}}%
\left( {\frac{\omega _{0h}}{{\tilde{\omega}}_{h}}}\right) ^{2}}}.
\end{equation}

The vectors $\mathbf{X},\mathbf{Y}$ and $\mathbf{Z}$ are unit vectors along
the corresponding Cartesian axes.

\begin{figure}
\includegraphics[]{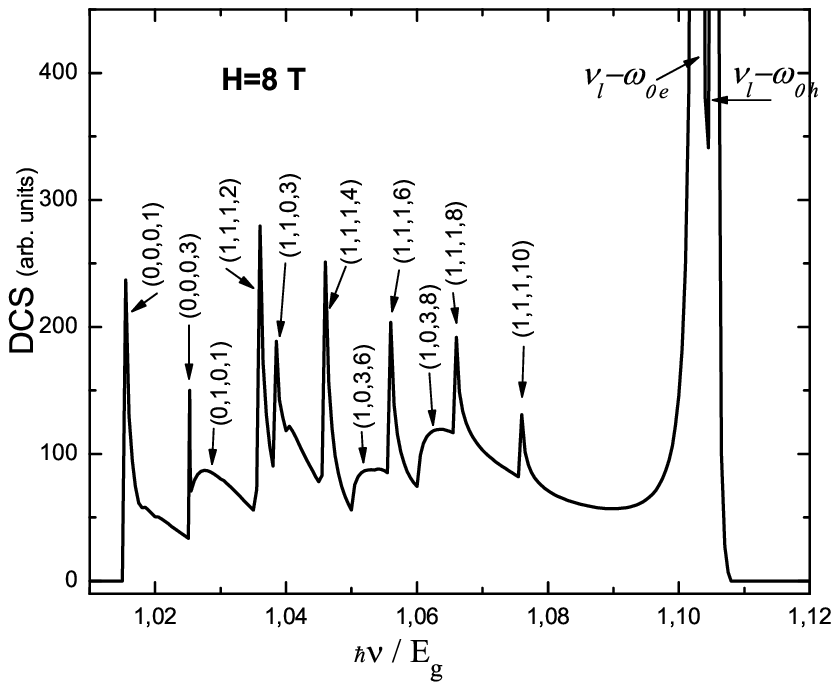}% Here is how to import EPS art
\caption[*]{\label{fg4.eps} The Raman spectra for the $Z$
scattering configuration for the magnetic fields  $H=8 T$. Other
parameters coincide with those of Figs. 1-3.}
\end{figure}

Let us make some remarks concerning the above equations. As indicated above,
when $\hbar \nu >E_{g}$ the Raman process involves the combination of
intraband and interband transitions. From Eq. (32) it follows that $%
J_{N_{2h},N_{2e}^{\prime }}$ ($J_{N_{2h}^{^{\prime }},N_{2e}}$) vanishes
unless $N_{2h}+N_{2e}^{\prime }=2n$ ($N_{2h}^{^{\prime }}+N_{2e}=2n$) where $%
n$ is an integer. So, transition can only take place between $N_{2h}$ and $%
N_{2e}^{\prime }$ ($N_{2h}^{^{\prime }}$ and $N_{2e}$) subbands with the
same parity ($2m\rightarrow 2n$ ; and $2m+1\rightarrow 2n+1$ ; $m$ and $n$
are integers). But for Eq. (31) quantum numbers $N_{1h}$ and $N_{1e}^{\prime
}$ ($N_{1h}^{^{\prime }}$ and $N_{1e}$) can change arbitrarily.

Hence, the following selection rules are obtained for interband transitions:
$$|N_{1h}-N_{1e}^{\prime }| =0,1,2,\ldots
;~~~|N_{2h}-N_{2e}^{\prime }|=0,2,4\ldots ; $$
$$|N_{1h}^{^{\prime }}-N_{1e}| =0,1,2,\ldots ;~~~|N_{2h}^{^{\prime
}}-N_{2e}|=0,2,4\ldots ;$$

1. When $H=0,~~x_{0e}=x_{0h}=0,~~{\tilde{L}}_{e}=L_{e}$ and ${\tilde{L}}%
_{h}=L_{h}$ that is the oscillator center of the conduction and valence band
electrons coincident. Then, making the replacement $N_{2h}\rightarrow
N_{1h},N_{2e}^{\prime }\rightarrow N_{1e}^{\prime }$ ($N_{2h}^{^{\prime
}}\rightarrow N_{1h}^{^{\prime }},N_{2e}\rightarrow N_{1e}$), we see that
Eq.(31) turns into Eq.(32). Thus, for $H=0$ we have the selection rules
$$
|N_{1h}-N_{1e}^{\prime }| =0,2,4\ldots ;|N_{2h}-N_{2e}^{\prime
}|=0,2,4\ldots ; $$
$$
 |N_{1h}^{^{\prime }}-N_{1e}| =0,2,4\ldots
;|N_{2h}^{^{\prime }}-N_{2e}|=0,2,4\ldots .
$$

2. The case $\omega _{e(h)}>>\omega _{0e(h)}$. This condition corresponds to
the strong magnetic fields, and we have
$$
{\tilde{\omega}}_{e(h)}=\omega _{e(h)}\sqrt{1+\left( {\frac{\omega _{0e(h)}}{%
\omega _{e(h)}}}\right) ^{2}}\approx \omega _{e(h)}
$$
and
$$
x_{0e}\approx x_{oh},~~{\tilde{L}}_{e}\approx {\tilde{L}}_{h}=\sqrt{{\frac{%
c\hbar }{eH}}}=l_{H},
$$
where $l_{H}$ is the magnetic length. For this case Eq. (31) differs from
zero when $N_{1h}=N_{1e}^{\prime }$ ($N_{1h}^{^{\prime }}=N_{1e}$). In this
way, the selection rule $|N_{2h}-N_{2e}^{\prime }|=2n+1$ ($|N_{2h}^{^{\prime
}}-N_{2e}|=2n+1$) takes place in intermediate magnetic fields.
\begin{figure}
\includegraphics[]{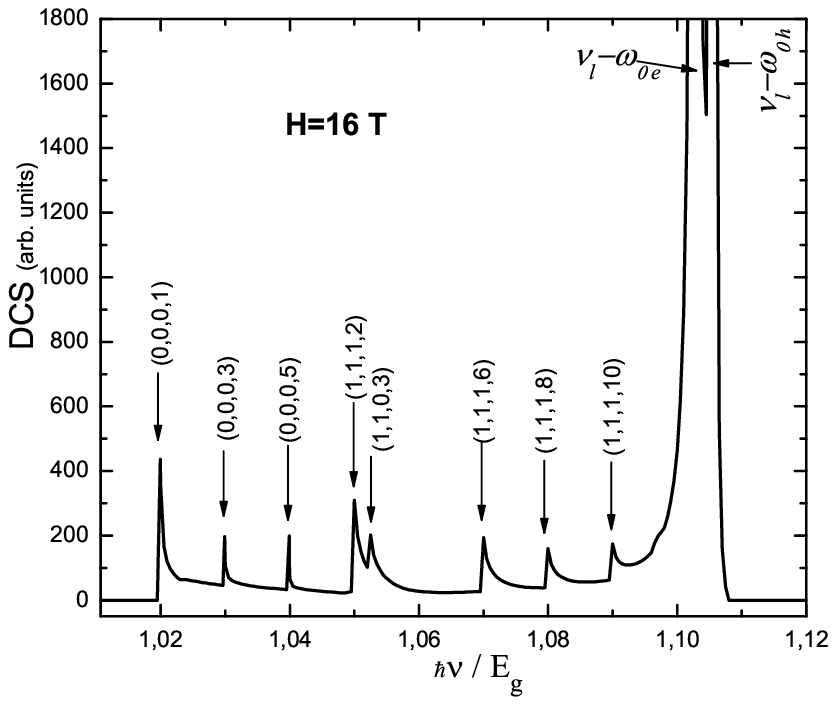}% Here is how to import EPS art
\caption[*]{\label{fg5.eps}The Raman spectra for the $Z$
scattering configuration for the magnetic fields  $H=16 T$. Other
parameters coincide with those of Figs. 1-3.}
\end{figure}

As can be seen from Eqs. (34) and (36) the differenial cross section is
directly proportional to the density-of -states of carriers in the valence
and conduction bands and to the interband matrix elements. In this case the
scattering spectrum shows density-of-states peaks and interband matrix
elements maximums. The positions of these structures are given as follows:
\begin{equation}
\label{42} \hbar \nu
=E_{N_{1h}}+E_{N_{2h}}+E_{N_{1e}}+E_{N_{2e}}+E_{g}.
\end{equation}
Here, the following selection rules must be fulfilled: $N_{1e}^{\prime
}=N_{1e}\pm 1,N_{2e}^{\prime }=N_{2e}$ ($N_{1h}^{\prime }=N_{1h}\pm
1,N_{2h}^{\prime }=N_{2h}$) for $X$ scattering configuration and $%
N_{2e}^{\prime }=N_{2e}\pm 1,N_{1e}^{\prime }=N_{1e}$ ($N_{2h}^{\prime
}=N_{2h}\pm 1,N_{1h}^{\prime }=N_{1h}$) for $Z$ scattering configuration. In
this case when $|N_{1h}-N_{1e}^{\prime }|=2n+1$ ($|N_{1h}^{^{\prime
}}-N_{1e}|=2n+1$) the spectrum shows maximums and when $|N_{1h}-N_{1e}^{%
\prime }|=2n$ ($|N_{1h}^{^{\prime }}-N_{1e}|=2n$) the ERS spectrum shows
singular peaks. The peaks and maximums related to these structures
correspond to interband EHP transitions and their positions depend on the
magnetic field.

Other singularities of equations (34) and (36) occur whenever $A(\nu )=0$
and $B(\nu )=0$. In the $X$ scattering configuration this singularities are
\begin{equation}
\label{43} \nu =\nu _{l}-{\tilde{\omega}}_{e},\ \ \ \nu =\nu
_{l}-{\tilde{\omega}}_{h}.
\end{equation}
Here the following selection rules are fulfilled: $N_{1e}^{\prime
}=N_{1e}+1,N_{2e}^{\prime }=N_{2e}$ and $N_{1h}^{\prime }=N_{1h}-1$, $%
N_{2h}^{\prime }=N_{2h}$.

For the $Z$ scattering configuration the Raman singularity is
\begin{equation}
\label{44} \nu =\nu _{l}-\omega _{0e},\ \nu =\nu _{l}-\omega
_{0h}.
\end{equation}
In this case the selection rules are: $N_{1e}^{\prime
}=N_{1e},N_{2e}^{\prime }=N_{2e}+1$ and $N_{1h}^{\prime }=N_{1h}$,
$ N_{2h}^{\prime }=N_{2h}-1$.

As can be seen from equations (43) and (44) these frequencies correspond to
electron transitions connecting the subband edges for a process involving
the conduction and valence bands (i.e., intraband transitions). We can also
notice that the $Y$ scattering configuration is free from Raman singularity
and relates to selection rules: $N_{1e}^{\prime }=N_{1e},N_{2e}^{\prime
}=N_{2e}$ and $N_{1h}^{\prime }=N_{1h}$, $N_{2h}^{\prime }=N_{2h}$

\section{Results and discussion}

In the following we present detailed numerical calculations of the
differential cross section of a GaAs/AlGaAs parabolic quantum wire
in presence of an uniform magnetic field as a function of $\hbar
\nu /E_{g}$ . The physical parameters used in our expressions
are:$ E_{g}=1.5177eV,~~m_{e}=0.0665m_{0},~~m_{h}=0.45m_{0}$ (the
heave-hole band). Taking the ratio 60:40 for the band-edge
discontinuity [20, 21], the conduction and valence barrier heights
are taken to be $\Delta _{e}=255 meV$ and $\Delta _{h}=170 meV$.
The oscillation frequencies $\omega _{0e}$ and $ \omega _{0h}$ of
the parabolic quantum wire are determined as
$$\omega _{0e(h)}={\frac{2}{d}}\sqrt{{\frac{2\Delta _{e(h)}}{m_{e(h)}}}},$$
where $d$ is the quantum wire diameter.

In Figs.1-3 we show the Raman spectra of the parabolic quantum wire in the $%
X $ scattering configuration for different magnetic fields. The diameter $d$
of the quantum wire is 2000 A. The incident radiation frequency $\hbar \nu
_{l}=1.68$ eV. The positions of the singularities are defined by Eqs.
(42),(43) and (44).

Figs. 3, 4 show the Raman spectra for the $Z$ scattering
configuration for the magnetic fields $H=8T$ and $H=16T$. Other
parameters coincide with those of Figs. 1-3. The structure of the
differential cross section, as given in Figs. provides a
transparent understanding of the energy subband structure of the
parabolic quantum wire in a transverse magnetic field.

In the present work we have applied a simplified model for the electronic
structure of the system. In a more realistic case we should consider the
real band structure using a calculation model like that of the
Luttinger-Kohn or Kane model. The above mentioned assumptions would lead to
better results, but entail more complicated calculations. However, within
the limits of our simple model we are able to take into account the
essential physical properties of the discussed problem. The fundamental
features of the differential cross section, as described in our work, should
not change very much in real quantum wire case.

It can be easily proved that the singular peak in the differential cross
section will be present irrespective of the model used for the subband
structure and may be determined for the values of $\hbar \nu _{s}$ equal to
the energy difference between two subbands $\hbar \nu _{s}=\hbar \nu
_{l}-\hbar \nu =E_{\alpha }^{e}-E_{\beta }^{e},$ where $E_{\alpha
}^{e}>E_{\beta }^{e}$ are electron energies in the subbands, respectively.
At present there is a lack of experimental work on this type of the ERS. Our
major aim in performing these calculations is to stimulate experimental
research in this direction.

\end{document}